# Clarification of Basic Concepts for Electromagnetic Interference Shielding Effectiveness

Mengyue Peng, Faxiang Qin[a]

*Institute for Composite Science Innovation (InCSI), School of Materials Science and Engineering, Zhejiang University, Hangzhou, 310027, China*

---

[a] Corresponding author: faxiangqin@zju.edu.cn

# ABSTRACT

There exists serious miscomprehension in the open literature about the electromagnetic interference shielding effectiveness (EMI SE) as a critical index to evaluate the shielding performance, which is misleading to the graduates and newcomers embarking on the field of electromagnetic shielding materials. EMI SE is defined as the sum of three terms including reflection loss, absorption loss and multiple reflection loss in the classical *Schelkunoff theory*, while it is decomposed into two terms named reflection loss and absorption loss in practice, which is called *Calculation theory* here. In this paper, we elucidate the widely-seen misconceptions connected with EMI SE via theoretical derivation and instance analysis. Firstly, the terms in *Calculation theory* are often mistakenly regarded as the approximation of the terms with the same names in *Schelkunoff theory* when multiple reflection loss is negligible. Secondly, it is insufficient and unreasonable to determine the absorption-dominant shielding performance in the case that absorption loss is higher than reflection loss since reflection loss and absorption loss cannot represent the actual levels of reflected and absorbed power. Power coefficients are recommended to compare the contribution of reflection and absorption to shielding performance. Thirdly, multiple reflection effect is included in the definitions of reflection loss and absorption loss in *Calculation theory*, and the effect of multiple reflections on shielding property is clarified as against the commonly wrong understandings. These clarifications offer correct comprehension about the shielding mechanism and assessment of reflection and absorption contribution to the total shielding.

# I. INTRODUCTION

The rapid advancements of telecommunication and proliferation of mobile electronic devices entail the development of high-performance electromagnetic interference (EMI) shielding materials to ensure the normal operation of equipment and protect humans from electromagnetic (EM) radiation pollution.[1-6] In general, EMI shielding materials can be classified into reflection-dominant and absorption-dominant types according to the reflection and absorption contribution to the total shielding. Conventional metal-based materials with remarkable electric conductivity have high reflection loss of the EM wave, enabling them as representative reflection-dominant shielding materials to block EM radiation. However, the majority of the EM wave is reflected for general high-conductive materials which results in secondary electromagnetic pollution.[7-9] Therefore, absorption-dominant shielding materials have become more desirable alternatives for EMI shielding applications.[8,10-13]

Fundamentally, it is primary and crucial to evaluate the EMI shielding performance and quantify the reflection and absorption contribution to shielding correctly for analyzing the shielding mechanism and developing absorption-dominant shielding materials. EMI shielding refers to the attenuation of the propagating EM wave produced by the shielding material. EMI shielding effectiveness (SE) is an important index to quantitatively evaluate the shielding performance. It is defined as the logarithm of the ratio, expressed in decibels, of transmitted power when there is no shield to the transmitted power when there is a shield.[14,15] The higher EMI SE, the less EM wave is transmitted through the shielding material.

EMI SE is defined as the sum of three terms including reflection loss ($SE_R$), absorption loss ($SE_A$) and multiple reflection loss ($SE_M$) in the most classical shielding theory originally developed by Schelkunoff,[16,17] while EMI SE is decomposed into two terms named reflection loss ($SE_R{'}$) and absorption loss ($SE_A{'}$) in practice.[13,18-20] The former and latter theories of EMI SE are called "*Schelkunoff theory*" and "*Calculation theory*" respectively in this paper for the sake of distinction. Confusion arises from that the terms in *Calculation theory* often use the same names or notations as the terms in *Schelkunoff theory*. The terms in *Calculation theory* are mistakenly regarded as the approximation of the terms in *Schelkunoff theory* when multiple reflection loss is negligible.[6,9,20-27] In fact, they are absolutely different physical quantities. Although the reflection loss and absorption loss describe reflection and absorption in both *Schelkunoff theory* and *Calculation theory*, none of them represent actual levels of reflected and absorbed power. Many publications incorrectly determine the shielding types of materials by comparing the contribution of reflection loss ($SE_R{'}$) and absorption loss ($SE_A{'}$) to the overall shielding effectiveness.[3,5,10,13,18,21,24,26,28-30]

Hence, some concepts connected with EMI SE are in urgent need of clarification. The motivation of this paper is to elucidate the widely existing misconceptions and suggest appropriate comprehension about EMI SE. The *Schelkunoff theory* and *Calculation theory* are elaborated to explain correlative terms about the reflection and absorption after depicting the interaction of EM wave with shielding materials. Then, the comparison between these two theories is implemented to demonstrate that the terms with the same names are actually different quantities. Finally, the contributions of reflection and absorption to the total shielding performance and the appropriate method of determining their contributions are elaborated.

## II. INTERACTION OF THE ELECTROMAGNETIC WAVE WITH A SHIELDING MATERIAL

When the EM wave is incident on the front interface between a shielding material and free space, the EM wave is reflected arising from the impedance mismatch between two different media. The un-reflected wave propagates towards the interior of the material and its strength exponentially decreases due to absorption. Once the wave reaches the rear interface, a portion of the wave passes through the shielding material and the rest portion is reflected back, generating an infinite number of multiple reflections between two interfaces. The schematic diagram of the interaction of the EM wave with a shielding material is depicted in Fig. 1.

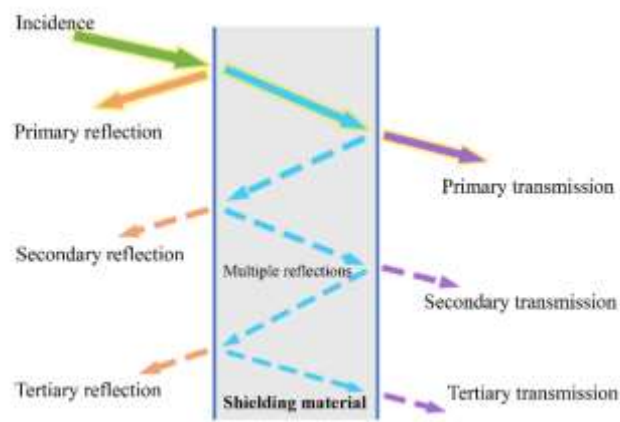

**Fig. 1.** Schematic of propagation of the EM wave through a material sheet.

## III. *SCHELKUNOFF THEORY* OF SHIELDING EFFECTIVENESS

Normally incident plane wave striking on a homogenous and isotropic material sheet is considered here for simplification, which is in accordance with the *Schelkunoff theory* based on transmission line model of shielding materials and ordinary SE measurement of materials. According to *Schelkunoff theory*, SE, generally expressed in logarithmic scale with the unit of decibel, is the sum of reflection loss ($SE_R$), absorption loss ($SE_A$) and multiple reflection loss ($SE_M$).[14,16,31,32]

$$\text{SE (dB)} = 10\log_{10}\left(\frac{P_i}{P_t}\right) = 20\log_{10}\left|\frac{E_i}{E_t}\right| = 20\log_{10}\left|\frac{H_i}{H_t}\right| = SE_R + SE_A + SE_M \quad (1)$$

where $P_i$ ($E_i$ or $H_i$) and $P_t$ ($E_t$ or $H_t$) are the power (intensity of the electric or magnetic field) of the incident and transmitted wave, respectively.

In the transmission line theory,[33] the characteristic impedance ($\eta$) of a medium is defined as

$$\eta = \sqrt{\frac{j\omega\mu}{\sigma + j\omega\varepsilon}} \quad (2)$$

where $\mu$, $\sigma$, and $\varepsilon$ are the permeability, conductivity and permittivity respectively; $\omega = 2\pi f$ is the angular frequency. The characteristic impedance of free space $Z_W \approx \sqrt{\frac{\mu_0}{\varepsilon_0}} = 377\ \Omega$, where $\mu_0$ and $\varepsilon_0$ are the permeability and permittivity of vacuum, respectively. It is customary to define the propagation constant ($\gamma$) in the medium such that

$$\gamma = (\alpha + j\beta) = \sqrt{j\omega\mu(\sigma + j\omega\varepsilon)} \quad (3)$$

where $\alpha$ is the attenuation constant and $\beta$ is the phase constant.

Eq. (2) and Eq. (3) are the general formulae of the characteristic impedance and propagation constant, respectively. For good conductors at the frequency where $\sigma \gg \omega\varepsilon$, the characteristic impedance and propagation constant are approximated as $\eta \approx \sqrt{\frac{j\omega\mu}{\sigma}}$ and $\gamma \approx \sqrt{j\omega\mu\sigma}$. For many materials like dielectric media and semiconductors, the characteristic impedance and propagation constant are approximated as $\eta \approx \sqrt{\frac{\mu}{\varepsilon}}$ and $\gamma \approx j\omega\sqrt{\mu\varepsilon}$, respectively. It is worth mentioning that both permeability and permittivity for materials with poor conductivity are complex values and the contribution of conduction loss is included in the dielectric loss.

**A. Reflection loss**

The reflection loss at an interface is related to the difference in the characteristic impedances of two media, as shown in Fig. 2.[9,14] This neglects the contribution of the multiple reflections between the two interfaces and absorption in the shielding material to shielding performance, merely considering the reduction in transmitted wave due to the primary reflection. The intensity of the reflected and transmitted wave from a medium with impedance $Z_1$ to a medium with impedance $Z_2$ can be expressed by

$$E_r = \Gamma E_i = \frac{Z_2 - Z_1}{Z_1 + Z_2} E_i \tag{4}$$

$$E_t = T E_i = \frac{2Z_2}{Z_1 + Z_2} E_i \tag{5}$$

where $E_i$, $E_r$ and $E_t$ are the intensity of the incident, reflected and transmitted wave, respectively; $\Gamma$ and $T = 1 + \Gamma$ are the reflection coefficient and transmission coefficient when the media are infinite in the thickness.

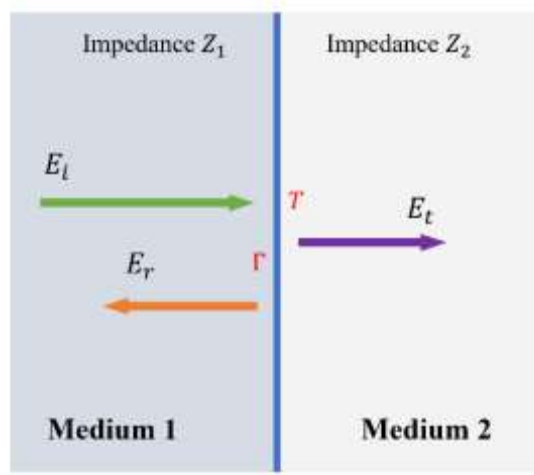

**Fig. 2.** Reflection and transmission of the incident wave at an interface between two infinite-thickness media.

When the EM wave passes through a shielding material, it encounters two interfaces between the material and free space, as shown in Fig. 3. The reflection coefficients and transmission coefficients at the fore and rear boundaries are $\Gamma_1 = \frac{Z_2-Z_1}{Z_1+Z_2}$, $\Gamma_2 = -\Gamma_1$, $T_{21} = 1 + \Gamma_1$, $T_{12} = 1 + \Gamma_2$, respectively. The transmitted wave ($E_t$) through the second interface is given by

$$E_t = T_{12}E_1 = T_{21}T_{12}E_i \tag{6}$$

According to the definition of shielding effectiveness, reflection loss is defined as

$$SE_R = 20\log_{10}\left|\frac{E_i}{E_t}\right| = 20\log_{10}\left|\frac{1}{T_{21}T_{12}}\right| \tag{7}$$

The reflection loss is equal to the reciprocal of the product of the transmission coefficients at both interfaces, expressed in decibel, indicating that reflection loss only depends on the impedance values of the media.

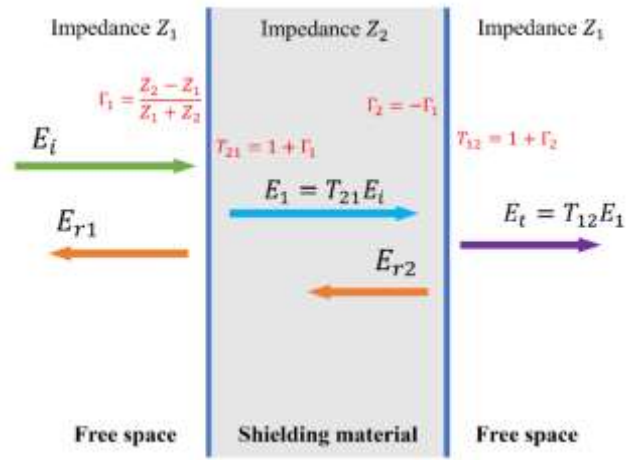

**Fig. 3.** Partial reflection and transmission occur at both faces of a shielding material.

**B. Absorption loss**

When the EM wave passes through a material, its amplitude decreases exponentially. This decay occurs ascribed to the dielectric loss, magnetic loss and conduction loss of the material.[14,34] Therefore, the electric field $E$ at a distance ($t$)

within the material is

$$E = E_i e^{-\gamma t} \tag{8}$$

The distance required for the wave to be attenuated to 1/e of its original value is defined as the skin depth ($\delta_s$). The absorption loss for a material with thickness $t$ is expressed as

$$SE_A = 20\log_{10}\left|\frac{E_i}{E_t}\right| = 20\log_{10}|e^{\gamma t}| \tag{9}$$

The absorption loss is closely related to the constitutive parameters ($\mu$, $\sigma$ and $\varepsilon$) and thickness of the material. It can be physically interpreted as the attenuation of transmitted wave owing to the loss within the shielding material when the impedance of shielding material is simultaneously matched with impedance of the surrounding and hence no reflection at both interfaces.

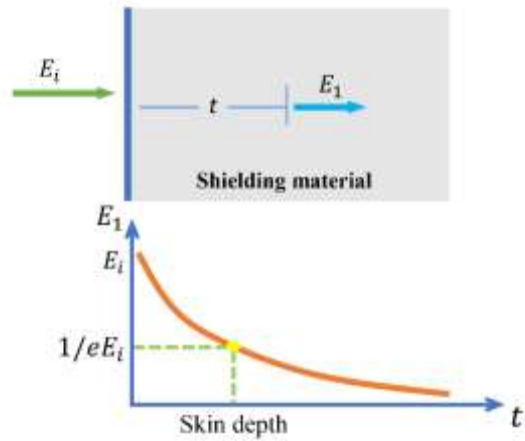

**Fig. 4.** Electromagnetic wave passing through an absorbing medium is attenuated exponentially.

**C. Multiple reflection loss**

If the shielding material is thin, the reflected wave from the second interface is re-reflected off the first interface, and then it returns to the second boundary to be reflected again. An infinite number of reflected waves occur, as shown in Fig. 5.[6] Therefore, not only reflection and absorption but also multiple reflections contribute to the overall shielding. The total transmitted (reflected) wave is the sum of all the partial waves transmitted (reflected) past the second (first) interface toward the right (left).[31,35]

$$E_t = E_{t1} + E_{t2} + \cdots + E_{tN} = T_{21}T_{12}e^{-\gamma t}E_i + T_{21}T_{12}e^{-3\gamma t}\Gamma_2^2 E_i + \cdots +$$
$$T_{21}T_{12}e^{-(2N-1)\gamma t}\Gamma_2^{(2N-2)}E_i = \left(\frac{T_{21}T_{12}e^{-\gamma t}}{1-\Gamma_1^2 e^{-2\gamma t}}\right)E_i \tag{10}$$

$$E_r = E_{r1} + E_{r2} + \cdots + E_{rN} = \Gamma_1 E_i + T_{21}T_{12}e^{-2\gamma t}\Gamma_2 E_i + \cdots +$$
$$T_{21}T_{12}e^{-(2N-2)\gamma t}\Gamma_2^{(2N-3)}E_i = \Gamma_1\left(1 - \frac{T_{21}T_{12}e^{-2\gamma t}}{1-\Gamma_1^2 e^{-2\gamma t}}\right)E_i \tag{11}$$

Hence, the corresponding power coefficients of transmissivity ($T$) and reflectivity ($R$) are

$$T = \frac{P_t}{P_i} = \left|\frac{E_t}{E_i}\right|^2 = \left|\frac{T_{21}T_{12}e^{-\gamma t}}{1-\Gamma_1^2 e^{-2\gamma t}}\right|^2 \tag{12}$$

$$R = \frac{P_r}{P_i} = \left|\frac{E_r}{E_i}\right|^2 = \left|\Gamma_1\left(1 - \frac{T_{21}T_{12}e^{-2\gamma t}}{1-\Gamma_1^2 e^{-2\gamma t}}\right)\right|^2 \tag{13}$$

By definition, SE is given by

$$SE \text{ (dB)} = 20\log_{10}\left|\frac{E_i}{E_t}\right| = 20\log_{10}\left|\frac{1-\Gamma_1^2 e^{-2\gamma t}}{T_{21}T_{12}e^{-\gamma t}}\right|$$
$$= 20\log_{10}\left|\frac{1}{T_{21}T_{12}}\right| + 20\log_{10}|e^{\gamma t}| + 20\log_{10}|1-\Gamma_1^2 e^{-2\gamma t}|$$
$$= SE_R + SE_A + SE_M \tag{14}$$

where the Correction term due to the successive re-reflections (also called multiple reflection loss) is defined as

$$SE_M = 20\log_{10}|1 - \Gamma_1^2 e^{-2\gamma t}| \tag{15}$$

Multiple reflection loss is important for electrically thin material since re-reflected waves increase transmitted energy so that it reduces the value of SE.[36,37] Multiple reflection effect can be safely neglected ($SE_M \approx 0$) when material thickness is bigger than skin depth or absorption loss ($SE_A$) is higher than 10 dB since the amplitude of the EM wave firstly reaching the second interface is negligible.[6,38]

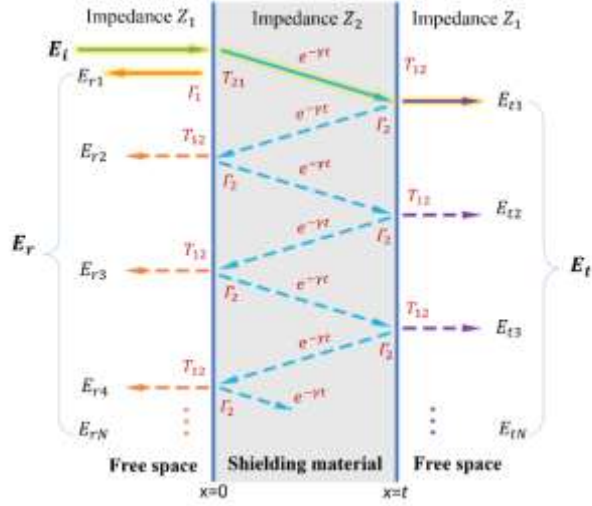

**Fig. 5.** Multiple reflection occurs in a material.

**D. Calculation of shielding effectiveness from scattering parameters**

Experimentally, the behavior of wave propagation through a sample can be recorded and measured by some technologies (e.g., Transmission/Reflection method) with different instruments (e.g., Vector Network Analyzer).[17,34,39,40] The reflected and transmitted waves in a two-port Vector Network Analyzer can be represented by the complex scattering parameters (S-parameters) i.e. $S_{11}$ ($S_{22}$) and $S_{21}$ ($S_{12}$) respectively, which in turn are correlated with reflectivity ($R$) and transmissivity ($T$).

$$R = |\frac{E_r}{E_i}|^2 = |S_{11}|^2 = |S_{22}|^2 \tag{16}$$

$$T = |\frac{E_t}{E_i}|^2 = |S_{21}|^2 = |S_{12}|^2 \tag{17}$$

The reflection coefficient $\Gamma_1$ ($|\Gamma_1| \leq 1$) at the fore interface of material is given by

$$\Gamma_1 = \chi \pm \sqrt{\chi^2 - 1} \tag{18}$$

where

$$\chi = \frac{S_{11}^2 - S_{21}^2 + 1}{2S_{11}} \tag{19}$$

The propagation factor $e^{-\gamma t}$ can be found by

$$e^{-\gamma t} = \frac{S_{11}+S_{21}-\Gamma_1}{1-(S_{11}+S_{21})\Gamma_1} \tag{20}$$

Then, the reflection loss, absorption loss and multiple reflection loss can be calculated from Eq. (18) - (20).

## IV. *CALCULATION THEORY* OF SHIELDING EFFECTIVENESS

From the perspective of electromagnetic energy, when the EM wave is incident on a material, the incident power is divided into the reflected power, absorbed power and transmitted power, as shown in Fig. 6.[29,41] The corresponding power coefficients of reflectivity ($R$), absorptivity ($A$) and transmissivity ($T$) follow the law of power balance ($R + A + T = 1$). Hence, the high shielding performance (low transmission of EM energy) results from two components, one associated with the high reflection of energy (defined as reflection loss ($SE_R{'}$)), and the other with high absorption of energy (defined as absorption loss ($SE_A{'}$)).[41-44]

$$SE = 10\log_{10}\left(\frac{P_i}{P_t}\right) = 10\log_{10}\left|\frac{1}{T}\right| = SE_R{'} + SE_A{'} \tag{21}$$

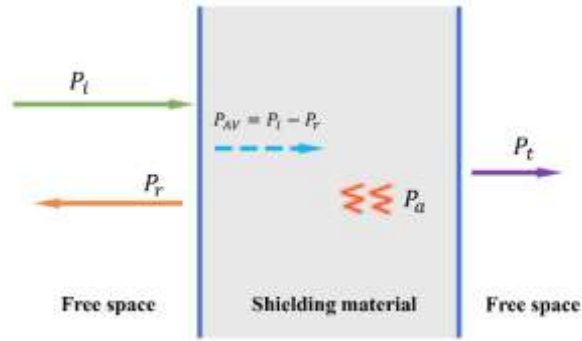

**Fig. 6.** Power distribution of total incident, reflected and transmitted wave through a material.

The EM energy incident at the fore interface of shielding material will be partially reflected from and partially transmitted into the sample where the later refers to here as "available power" ($P_{AV} = P_i - P_r$) which designates the net power available for entering the material. The reflection loss ($SE_R'$) is defined by

$$SE_R' = 10\log_{10}\left(\frac{P_i}{P_{AV}}\right) = 10\log_{10}\left|\frac{1}{1-R}\right| = 10\log_{10}\left|\frac{1}{1-S_{11}^2}\right| \qquad (22)$$

$SE_R'$ is the ratio of the incident power to available power, representing the reduction in the incident wave traveling into the material resulting from reflection.

The available power is then absorbed by the material while the remaining power is transmitted across the rear interface. The absorption loss ($SE_A'$) is defined as

$$SE_A' = 10\log_{10}\left(\frac{P_{AV}}{P_t}\right) = 10\log_{10}\left|\frac{1-R}{T}\right| = 10\log_{10}\left|\frac{1-S_{11}^2}{S_{21}^2}\right| \qquad (23)$$

$SE_A'$ is the ratio of available power to transmitted power, and represents the ability of the material to attenuate EM wave that only penetrates the material due to absorption.[37,45,46]

## V. CLARIFICATION OF THE SHIELDING EFFECTIVENESS CONCEPTS

**A. Comparison of the terms in *Schelkunoff theory* and *Calculation theory***

Although reflection loss (absorption loss) exists in both *Schelkunoff theory* and *Calculation theory*, they are totally different physical quantities. As mentioned above, it is apparent that reflection loss (absorption loss) in both theories possess absolutely different physical meanings. The comparison of the values of reflection loss (absorption loss) is exhibited in Appendix A. The formulas of $SE_R$ and $SE_R'$ are different and $SE_R$ is independent of the material thickness while $SE_R'$ is not. It is evident that $SE_A$ is unequal to $SE_A'$ by comparison of Eq. (A2) and (A5). Hence, the reflection loss (absorption loss) in *Schelkunoff theory* is not equal to the reflection loss (absorption loss) in *Calculation theory*.

To further prove this point, the terms in *Schelkunoff theory* and *Calculation theory* for different materials are compared in Fig. 7. For an electrically thick material with high conductivity from 2.5 kHz to 25 GHz in Fig. 7(a), the values of corresponding quantities with the same name are not equivalent apparently. $SE_M$ approaches to zero, indicating multiple reflection loss can be ignored. $SE_R'$ value is approximately equal to the half of $SE_R$ while $SE_A'$ is greater than $SE_A$, which is consistent with the theoretical calculation for the high-conductivity and thick material in Appendix A. More importantly, it contradicts the common opinion that reflection loss ($SE_R'$) and absorption loss ($SE_A'$) in *Calculation theory* are approximations for the reflection loss ($SE_R$) and absorption loss ($SE_A$) in *Schelkunoff theory* when multiple reflection loss is negligible. For a high-conductivity material with a small thickness in Fig. 7(b), $SE_M$ has a relatively large negative value at lower frequencies, contributing negatively to the total shielding effectiveness. EMI SE of a material with low conductivity is shown in Fig. 7(c), the value of $SE_R$ is bigger than $SE_R'$ while $SE_A$ is smaller than $SE_A'$ at the majority of the frequencies. However, $SE_R$ ($SE_A$) is approximately equal to $SE_R'$ ($SE_A'$) at higher frequencies above ~10 GHz. Specially, Fig. 7(d) exhibits a practical case of carbon black/polylactic acid (CB/PLA) composite material at the range from 2 GHz to 18 GHz (see Appendix B for more details about preparation, electromagnetic measurement and permittivity (Fig. 9) of the CB/PLA composite). The value of $SE_R$ ($SE_A$) is small and approximates $SE_R'$ ($SE_A'$). The fact that the terms with the same names are similar in values for some materials at certain frequencies often induces the misconceptions that they are the same physical quantities and the deviation in their values results from the small divergence between theoretical calculation and experimental measurement.[21,47]

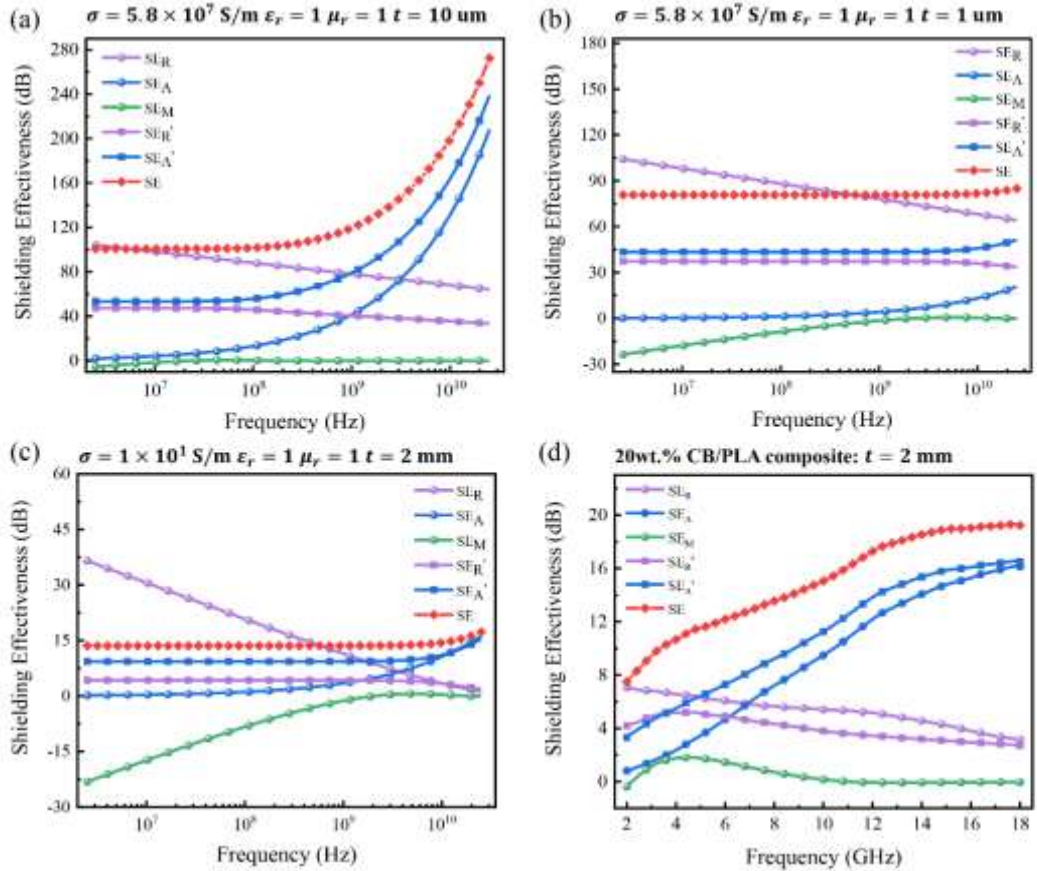

**Fig. 7.** Comparison of the terms in *Schelkunoff theory* and *Calculation theory*; (a) a material with σ = 5.8 × 10$^7$ S/m and t = 10 um; (b) a material with σ = 5.8 × 10$^7$ S/m and t = 1 um; (c) a material with σ = 1 × 10$^1$ S/m and t = 2 mm; (d) 20wt.% CB/PLA composite material with t = 2 mm.

If the dimensions of microstructures and inhomogeneities of materials are much smaller than the wavelength of the incident wave (e.g., micro-/nano-composites at gigahertz), the shielding materials can be regarded as homogeneous media. Both *Schelknoff theory* and *Calculation theory* can be applied to analyze the shielding performance of these materials. However, it is more straightforward to calculate the terms from the measured S-parameters in *Calculation theory* than *Schelkunoff theory*. The model of *Schelkunoff theory* in this paper is suitable for single-layered materials. However, the simple model of *Calculation theory* has the same physical meanings for single-layered and multi-layered materials and can be applied to both of them. Moreover, the reflection loss and absorption loss are closely related to the reflected and absorbed power in *Calculation theory*. Hence, *Calculation theory* is often adopted to calculate EMI SE and analyze the shielding mechanism in practice.

**B. Contribution of the reflection and absorption to shielding performance**

Although the reflection loss and absorption loss describe reflection and absorption respectively in *Schelkunoff theory* as well as *Calculation theory*, none of them represent actual levels of reflected and absorbed power.[37,46] Many publications incorrectly determine the shielding mechanism by comparing the contribution of reflection loss $SE_R'$ and absorption loss $SE_A'$ to the overall EMI SE.[3,5,18,21,28,29,48] When $SE_A'$ is higher than $SE_R'$, it does not mean that the contribution of absorption is larger than that of reflection or absorption is the dominant shielding mechanism.[45,46,49] Take the shielding effectiveness of the CB/PLA composite material for an example. Although $SE_A'$ is bigger than $SE_R'$, $A$ is smaller than $R$ at the range from 4 GHz to 15 GHz, as shown in Fig. 8(a). More than 50% of the EM energy will be reflected when $SE_R'$ is higher than 3 dB according to the relationship of $R$ and $SE_R'$ shown in Fig. 8(b). The remaining energy of less than 50% is available to travel into the shielding material. Even if the material has a strong ability of attenuating or absorbing the electromagnetic wave penetrating into the shielding material which means the very large value of absorption loss ($SE_A'$), the contribution of absorption to the total shielding must be less than 50% according to the law of power balance. In this case, reflection is the dominant shielding mechanism rather than absorption. It is reasonable and intuitive to adopt power coefficients of $R$ and $A$ to determine the type of shielding materials and shielding mechanisms.[8,49] If $A$ is higher than $R$, absorption is the dominant shielding mechanism. The absorption contribution fraction to shielding can be defined as the ratio of $A$ to ($A+R$).

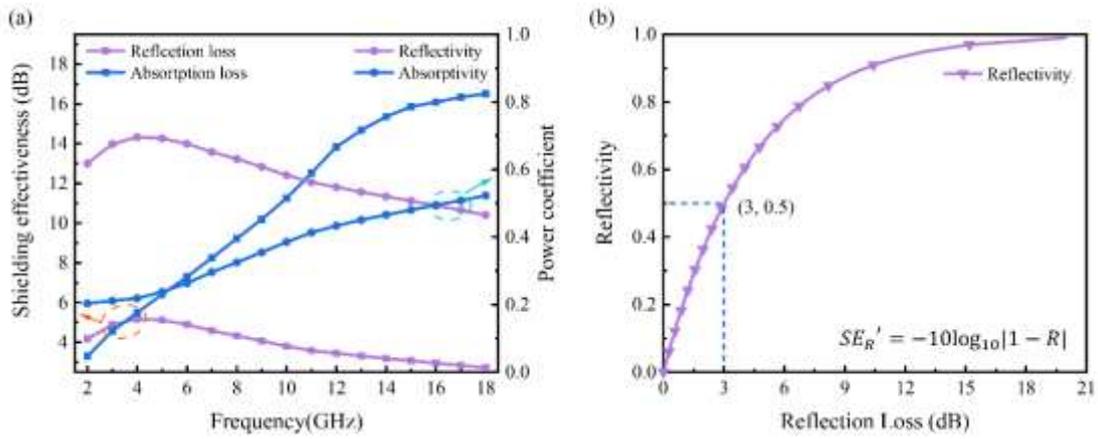

**Fig. 8.** Comparison of the reflection and absorption contribution to shielding. (a) Reflection loss, absorption loss, reflection and absorption of the CB/PLA composite material. (b) The curve of reflectivity changed with reflection loss.

Based on the inaccurate distinction between reflection and multiple reflections, multiple reflections are wrongly considered a mechanism that increases the overall shielding effectiveness.[37,50] Typically, multiple reflections decrease the shielding effectiveness since re-reflected waves between the front and back boundaries of the material increase the transmitted energy when the shielding material is thinner than skin depth. However, when the material thickness is larger than the skin depth or $SE_A \geq$ 10 dB, the effect of multiple reflection can be neglected.

In addition, the concept of multiple reflections between both interfaces between the shielding material and free space sometimes is confused with the concept of internal multiple reflections resulting from the abundant interfaces of porous materials.[3,9,51,52] The internal multiple reflections at the lots of microscopic interfaces inside the material prolong the transmission paths of the EM wave and further help absorb more EM wave, but it does not necessarily enhance the shielding performance.[4,52] These microscopic interactions of the material and EM wave can be included in the macroscopic effective constitutive parameters if the characteristic sizes of heterogeneities are much smaller than the wavelength in the material. The effective constitutive parameters have complex mathematical relationships with the shielding effectiveness. Hence, it is hard to determine the contribution from microscopic multiple-reflections to the shielding performance. Multiple reflections in shielding refer to the infinite number of reflected waves occurring at the fore and rear interfaces between the material and free space which is a concept at the macro level and is heavily dependent on the material thickness.

It is worth mentioning that multiple reflections are the behavior of the EM wave in the propagation process and will finally be embodied in the reflected, absorbed and transmitted EM energy from the perspective of energy. Hence when *Calculation theory* is employed to calculate EMI SE and analyze shielding mechanisms in practice, it is illogical to use the concept of multiple reflection loss since the effect of multiple reflections is included in the definition of reflection loss ($SE_R{'}$) and absorption loss ($SE_A{'}$).

# VI. CONCLUSION

Reflection loss (absorption loss) in *Schelkunoff theory* and *Calculation theory* are different physical quantities, respectively. Reflection loss (absorption loss) in *Calculation theory* cannot be regarded as the approximation of reflection loss (absorption loss) in *Schelkunoff theory* when multiple reflection loss is negligible. Reflection loss and absorption loss cannot represent the actual levels of reflected and absorbed power, hence it is unreasonable to determine the reflection and absorption contribution to the overall shielding by comparison of reflection loss and absorption loss. It is suggested to adopt power coefficients of reflectivity and absorptivity to describe the shielding mechanisms. Multiple reflection loss gives a negative contribution to shielding effectiveness for electrically thin materials while it can be negligible for thick materials. The concept of Multiple reflection loss does not exist in *Calculation theory* since the multiple reflection effect is included in the reflection loss and absorption loss. Only accurate comprehension of the concepts about shielding effectiveness can allow a proper understanding of shielding mechanisms and the successful development of high-performance shielding materials.

# ACKNOWLEDGEMENT

This work is supported by ZJNSF No. LR20E010001 and National Key Research and Development Program of China No. 2021YFE0100500 and Zhejiang Provincial Key Research and Development Program (2021C01004) and Chao Kuang Piu High Tech Development Fund 2020ZL012 and Aeronautical Science Foundation 2019ZF076002.

## DATA AVAILABILITY

The data that support the findings of this study are available from the corresponding author upon reasonable request.

## APPENDIX A: CALCULATION OF THE TERMS IN *SCHELKUNOFF THEORY* AND *CALCULATION THEORY*

For comparing the values of terms in *Schelkunoff theory* and *Calculation theory*, the formulas of these terms are all represented by variables of reflection coefficient $\Gamma_1$ and propagation factor $e^{-\gamma t}$.

The terms in *Schelkunoff theory* are expressed by

$$SE_R = 20\log_{10}\left|\frac{1}{1-\Gamma_1^2}\right| \tag{A1}$$

$$SE_A = 20\log_{10}|e^{\gamma t}| \tag{A2}$$

$$SE_M = 20\log_{10}\left|1 - \Gamma_1^2 e^{-2\gamma t}\right| \tag{A3}$$

Combining Eq. (12), (13), (22) and (23), the terms in *Calculation theory* can be obtained

$$SE_R' = -10\log_{10}\left|1 - |\Gamma_1 \frac{1-e^{-2\gamma t}}{1-\Gamma_1^2 e^{-2\gamma t}}|^2\right| \tag{A4}$$

$$SE_A' = 20\log_{10}|e^{\gamma t}| + 10\log_{10}\left|1 - |\Gamma_1 \frac{1-e^{-2\gamma t}}{1-\Gamma_1^2 e^{-2\gamma t}}|^2\right| - 20\log_{10}\left|\frac{1-\Gamma_1^2}{1-\Gamma_1^2 e^{-2\gamma t}}\right| \tag{A5}$$

For the high-conductivity shielding material with a thickness much bigger than skin depth (t $\gg \delta_s = 1/\alpha$), the propagation constant can be approximated as $\gamma \approx (1+j)\alpha = (1+j)\sqrt{\pi f \mu \sigma}$. The value of $e^{-2\gamma t}$ is very small ($e^{-2\gamma t} \approx 0$). Multiple reflection loss can be neglected ($SE_M \approx 0$). The reflection loss and absorption loss in *Calculation theory* can be approximated as:

$$SE_R' \approx 10\log_{10}\left|\frac{1}{1-\Gamma_1^2}\right| \tag{A6}$$

$$SE_A' \approx 20\log_{10}|e^{\gamma t}| + 10\log_{10}\left|\frac{1}{1-\Gamma_1^2}\right| \tag{A7}$$

The value of $SE_R'$ is approximately equal to the half of $SE_R$ and $SE_A'$ is greater than $SE_A$.

# APPENDIX B: COMPLEX PERMITTIVITY OF THE CB/PLA COMPOSITE MATERIAL

The carbon black/polylactic acid (CB/PLA) composite with 20wt.% content of CB was prepared by extruding the mixture of CB and PLA particles (a 2:8 ratio by mass) via a twin-screw extruder. The composite powders were hot-pressed at 140℃ into a standard toroidal shape (inner diameter: 3.04 mm, outer diameter: 7mm, thickness: 2 mm) for the measurement of EM parameters. The scattering parameters of the composite were measured by a Vector Network Analyzer (R&S ZNB20) at the frequency range of 2-18 GHz with coaxial lines. The complex effective permittivity value of the composite (Fig. 9) was retrieved from the scattering parameters with the Nicolson-Ross-Weir (NRW) method.[34]

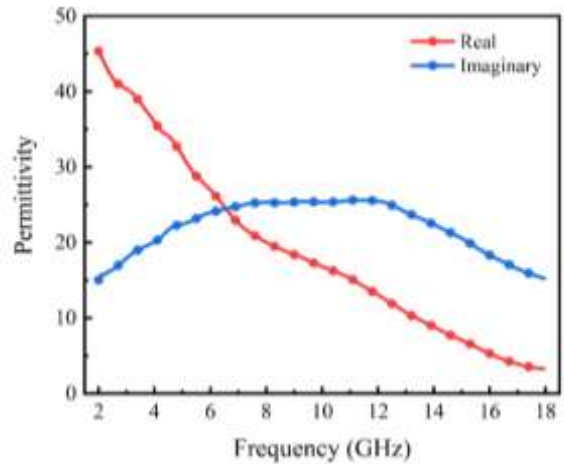

**Fig. 9.** Complex permittivity of the 20wt.% CB/PLA composite material.